# Phase diagrams and superconductivity of ternary Ca-Al-H compounds under high pressure

Ming Xu[1], Defang Duan[2], Wendi Zhao[1], Decheng An[2], Hao Song[1,*] and Tian Cui[1,2]†

[1]*Institute of High Pressure Physics, School of Physical Science and Technology, Ningbo University, Ningbo 315211, China*

[2]*State Key Laboratory of Superhard Materials, College of Physics, Jilin University, Changchun 130012, China*

*Corresponding author: songhao@nbu.edu.cn

†Corresponding author: cuitian@nbu.edu.cn

**Abstract**

The search for high-temperature superconductors in hydrides under high pressure has always been a research hotspot. Hydrogen-based superconductors offer an avenue to achieve the long-sought goal of superconductivity at room temperature. We systematically explore the high-pressure phase diagram, electronic properties, lattice dynamics and superconductivity of the ternary Ca-Al-H system using *ab initio* methods. We found two stable ternary hydrides at 50 GPa: *Cmcm*-$CaAlH_5$ and *Pnnm*-$CaAl_2H_8$, which both are semiconductors. At 200 GPa, a new phase of *P2$_1$/m*-$CaAlH_5$, *P4/mmm*-$CaAlH_7$ and a metastable compound *Immm*-$Ca_2AlH_{12}$ were found. Furthermore, *P4/mmm*-$CaAlH_7$ has obvious phonon softening of high frequency vibrations along the Z-A direction, point A and point X, which improves the strength of electron-phonon coupling. Therefore, a superconducting transition temperature $T_c$ of 71 K is generated at 50 GPa. In addition, the thermodynamic metastable *Immm*-$Ca_2AlH_{12}$ exhibits a superconducting transition temperature of 118 K at 250 GPa. These results are very useful for the experimental searching of new high-$T_c$ superconductors in ternary hydrides.

## I. INTRODUCTION

The researches on superconducting properties of materials have always been one of the hotspots in condensed matter physics, and the room temperature superconductivity has always been the long-sought goal pursued by most researchers. Since hydrogen is the lightest element in nature and exhibit substantially high Debye temperature[1], metallic hydrogen has been expected to be a candidate for room-temperature superconductor and synthesize under high pressure[2], so researchers have set off an upsurge in the investigation for metallic hydrogen for a long time. The metallization of hydrogen requires extremely high-pressure conditions[3-5], which is difficult to achieve. The experimental observation of metallic hydrogen is still controversial. But at the beginning of this century, Aschroft came up with a great pioneering idea[6], "chemical precompression", which means that hydrogen-rich materials probably can be metallized at lower pressures and exhibit high temperature superconductivity. So far, the most of binary hydrides have been explored theoretically, and many excellent results have been obtained[7-10]. The superconducting transition temperature $T_c$ exceeds 200 K in binary hydrides at high pressure, such as $HfH_{10}$ with $T_c$ of 234 K at 250 GPa[11], $YH_6$ with $T_c$ of 264 K at 120 GPa[12], $YH_{10}$ with $T_c$ of 303 K at 400 GPa[13], $ThH_{10}$ with $T_c$ of 241 K at 100 GPa[14], $CaH_6$ with $T_c$ of 235 K at 150 GPa[15].

Especially the theoretical and experimental results of LaH$_{10}$[16-19] and H$_3$S[20-23] have great significance in the field of high temperature superconductivity. H$_3$S exhibits a high $T_c$ of 203 K and adopting a covalent sixfold cubic structure. The hydrogen-rich compound LaH$_{10}$ has H$_{32}$ cage with a sodalite-like structure, showing a high $T_c$ of 250 K at 170 GPa. The common features[7] of these high-temperature hydrogen-rich superconductors are relatively high symmetry, high density of states at the Fermi level and strong electron-phonon coupling.

As the binary hydrides have been widely studied, the researches on ternary hydrides have gradually begun. Comparing with binary hydrides, ternary hydrides offer more abundant structures caused by the diversity of chemical compositions. Hydrogen-rich ternary compounds may undergo metallization at lower pressures due to the "chemical precompression" from the third element. Moreover, the hydrogen-rich compounds are more likely to exhibit high density of states at the Fermi level, which makes them strong contenders for high-temperature superconductors. Recently, the theoretical prediction on the ternary hydrides have also achieved good results[24-28]. For example, Li$_2$MgH$_{16}$ exhibits an unprecedentedly high $T_c$ of 473 K at 250 GPa[27]. Ternary compounds formed by introducing lithium into the yttrium hydrides under pressure, especially LiYH$_9$ with a $T_c$ of 109 K at 250 GPa[25]. Through theoretical calculations, CaBH$_7$ has a $T_c$ of 135 K at 150 GPa[24,28], and NaAlH$_8$ exhibits phonon anomalies which caused by Fermi surface nesting with a $T_c$ of 55 K at 300 GPa[26].

In this paper, we report a first-principles study results of the Ca-Al-H system. The *Pnnm*-CaAl$_2$H$_8$ and *Cmcm*-CaAlH$_5$ are semiconductors at 50 GPa. *P4/mmm*-CaAlH$_7$ is thermodynamic stable at 200 GPa, while the *Immm*-Ca$_2$AlH$_{12}$ is metastable. CaAlH$_7$ and Ca$_2$AlH$_{12}$ are good metals which both containing H and H$_2$ units, with superconducting transition temperatures of 71 K at 50 GPa and 118 K at 250 GPa, respectively. We proposed five possible synthetic routes (Ca+Al+H$_2$, Ca+AlH$_3$+H$_2$, CaH$_2$+Al+H$_2$, CaH$_2$+AlH$_3$+H$_2$, CaH$_4$+Al+H$_2$) and demonstrate their enthalpies of formation for the Ca-Al-H compounds, which will contribute to subsequent experimental studies. Our work has important implications for understanding the high-pressure phase diagram and fundamental properties of the Ca-Al-H ternary system.

## II. COMPUTATIONAL DETAILS

We performed variable composition structure searches using *ab initio* random structure searching code (AIRSS)[29,30]. For Ca$_x$Al$_y$H$_z$ (1≤$x$≤2, 1≤$y$≤2, 1≤$z$≤12) at 50 and 200 GPa over 10,000 structures were predicted. The Cambridge Series Total Energy Package (CASTEP)[31-33] software was used to determine the stable structure on the ternary phase diagram. The cut-off energy of the plane wave was set to 300 eV, and the sampling grid spacing of the selected Brillouin zone was 2π×0.07 Å$^{-1}$. the candidates and previously reported structures underwent high-quality parameter re-optimization. To precisely converge the enthalpy calculation to less than 1 meV/atom, the Brillouin zone *k*-point mesh was sampled with 2π×0.03 Å$^{-1}$, and the cut-off energy was set to be 800 eV.

Calculations of electronic properties were performed using the Vienna *ab initio* Simulation Package (VASP)[34], the projected augmented wave method (PAW)[35] was used to simulate the interaction between electrons and ions. 3$s^2$3$p^6$4$s^2$, 3$s^2$3$p^1$ and 1$s^1$ were considered as valence electrons for Ca, Al and H, respectively. We used a generalized gradient approximation (GGA) with the Perdew-Burke-Ernzerhof (PBE) parametrization for the exchange-correlation functional[36]. The cut-off energy was set to 800 eV and the Monkhorst-Pack $k$ mesh $2\pi \times 0.03$ Å$^{-1}$ was used, a denser $k$ mesh $2\pi \times 0.01$ Å$^{-1}$ was used to obtain high precise results of electronic properties. Phonon spectra and electron-phonon coupling were calculated using the Quantum-Espresso package[37]. The ultrasoft pseudopotential (USPP) and the plane wave cut-off energy of 80 Ry were used. For $P4/mmm$-CaAlH$_7$ 24×24×16 $k$ mesh and 6×6×4 $q$ mesh were used, for $Immm$-Ca$_2$AlH$_{12}$ 16×16×16 $k$ mesh and 4×4×4 $q$ mesh were used. Considering that the EPC parameters $\lambda$ of the all hydrides we investigated are less than 1.5, the Allen-Dynes modified McMillan equation[38] is selected to estimate the superconducting transition temperature $T_c$.

## III. RESULTS AND DISCUSSION
### A. Ternary phase diagrams and different synthetic routes of ternary hydrides

Ca-H and Al-H binary compounds have been systematically studied, and some binary compounds exhibit superconducting properties under high pressure[15,39,40]. Here, we performed structure searches for Ca$_x$Al$_y$H$_z$ (1≤$x$≤2, 1≤$y$≤2, 1≤$z$≤12) and constructed ternary Ca-Al-H phase diagrams at 50 and 200 GPa, respectively. As shown in Fig.1, the red solid and hollow circles represent stable and metastable ternary hydrides, respectively. We predicted several stable and metastable ternary hydrides. $Cmcm$-CaAlH$_5$ and $Pnnm$-CaAl$_2$H$_8$ are stable at 50 GPa; $P2_1/m$-CaAlH$_5$, and $P4/mmm$-CaAlH$_7$ are stable up to 200 GPa; $Immm$-Ca$_2$AlH$_{12}$ is metastable (31 meV/atom above the convex hull at 200 GPa), and this does not preclude Ca$_2$AlH$_{12}$ from experimental synthesis because many compounds observed in the high-pressure experiments were metastable[41-43]. The elements and binary hydrides that are stable under high pressures were referred from previous works[15,39,40,44].

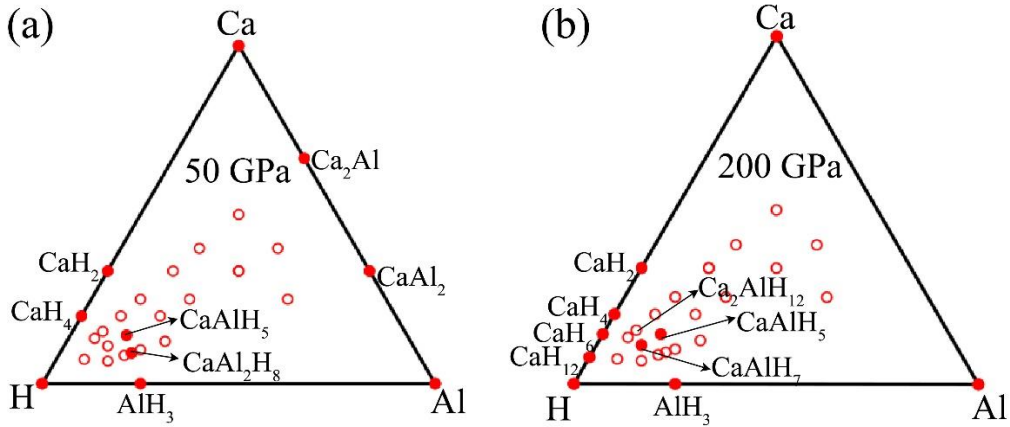

Fig.1. Ternary phase diagrams of the Ca-Al-H system at different pressures. Solid red circles indicate stable structures and open red circles indicate metastable structures.

The crystal structures of the $P4/mmm$-CaAlH$_7$ and $Immm$-Ca$_2$AlH$_{12}$ are shown in Fig. 2, and crystal structure of other hydrides can be found in Supplementary Material Fig. S1, and their detailed lattice parameters and atomic coordinates are shown in the Table S1 of the Supplementary Material. As can be seen from Fig. 2(c), CaAlH$_5$ is thermodynamic stable above 50 GPa, and it has two energy-competing structures. According to our calculations, at 50-90 GPa and 120-130 GPa, the enthalpy of the $Cmcm$ phase is lower than that of the $P2_1/m$. However, at 90-120 GPa and 130-200 GPa, the enthalpy of the $P2_1/m$ phase is lower than that of the $Cmcm$ phase and becomes more stable. CaAlH$_7$ has a thermodynamically stable interval of 80-200 GPa and the phase transform from *Cmcm* to *P4/mmm* occurs at 120 GPa.

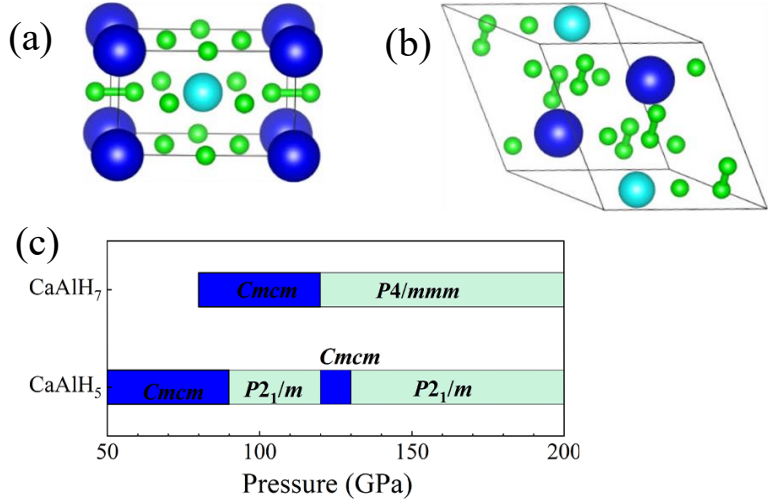

Fig.2. The crystal structures of (a) *P4/mmm*-CaAlH$_7$ and (b) *Immm*-Ca$_2$AlH$_{12}$. Calcium atoms are blue, aluminum atoms are cyan, and hydrogen atoms are green. (c) Predicted pressure-composition phase diagram of CaAlH$_5$ and CaAlH$_7$.

Because there are relatively abundant precursors, ternary Ca-Al-H compounds can be synthesized through different routes under high pressure. For example, Ca + Al + H$_2$, Ca + AlH$_3$ + H$_2$, CaH$_2$ + Al + H$_2$, CaH$_4$ + Al + H$_2$, CaH$_6$ + Al + H$_2$, CaH$_2$ + AlH$_3$ + H$_2$, CaH$_4$ + AlH$_3$ + H$_2$. Here we mainly focus on five possible synthetic routes (Ca+Al+H$_2$, Ca+AlH$_3$+H$_2$, CaH$_2$+Al+H$_2$, CaH$_2$+AlH$_3$+H$_2$, CaH$_4$+Al+H$_2$) and calculate their enthalpies of formation, as shown in Table I. Comparing the formation enthalpies of various synthesis routes, compressing the mixture of CaH$_2$+AlH$_3$+H$_2$ is the most favorable route to synthesize Ca$_x$Al$_y$H$_z$ (see Table I). From another perspective, CaAlH$_z$ ($z$ > 5) can also be synthesized by extruding CaAlH$_5$ + H$_2$. We believe that some other synthetic routes may be obtained by extending the precursors or changing the combinations. Once these synthetic pathways are identified, they can provide informative guidance for future experimental synthesis.

TABLE I. Formation enthalpies (eV/atom) of Ca$_x$Al$_y$H$_z$ at 50 and 200 GPa. $\Delta H_1$, $\Delta H_2$, $\Delta H_3$, $\Delta H_4$, $\Delta H_5$ represent the formation enthalpies of Ca$_x$Al$_y$H$_z$ from Ca + Al + H$_2$, Ca + AlH$_3$ + H$_2$, CaH$_2$ + Al + H$_2$, CaH$_2$ + AlH$_3$ + H$_2$, and CaH$_4$ + Al + H$_2$, respectively.

| | | $\Delta H_1$ | $\Delta H_2$ | $\Delta H_3$ | $\Delta H_4$ | $\Delta H_5$ |
|---|---|---|---|---|---|---|
| 50 GPa | *Cmcm*-CaAlH$_5$ | -0.788 | -0.676 | -0.226 | -0.115 | -0.178 |
| | *Pnnm*-CaAl$_2$H$_8$ | -0.581 | -0.438 | -0.223 | -0.081 | -0.190 |
| 200 GPa | *P2$_1$/m*-CaAlH$_5$ | -0.847 | -0.498 | -0.385 | -0.036 | -0.254 |
| | *P4/mmm*-CaAlH$_7$ | -0.765 | -0.493 | -0.405 | -0.134 | -0.303 |
| | *Immm*-Ca$_2$AlH$_{12}$ | -0.724 | -0.561 | -0.293 | -0.130 | -0.170 |

### B. Electronic properties of ternary hydrides

Electronic band structures and density of states (DOS) are used to analyze electronic properties. As shown in Fig.3, we found that *Cmcm*-CaAlH$_5$ and *Pnnm*-CaAl$_2$H$_8$ are indirect bandgap semiconductors at 50 GPa. *P2$_1$/m*-CaAlH$_5$ exhibit weak metallic states, and the DOS near the Fermi level is almost zero (see Fig. S2 in the Supplemental Material), so we infer that this structure do not have high temperature superconductivity. The conduction and valence bands of $P4/mmm$-$CaAlH_7$ and $Immm$-$Ca_2AlH_{12}$ overlap at the Fermi level, and the electronic density of states at the Fermi level is dominated by the electrons from H, as shown in Fig.3, which means that these two structures may have better superconducting properties. Then, we mainly discuss these two structures.

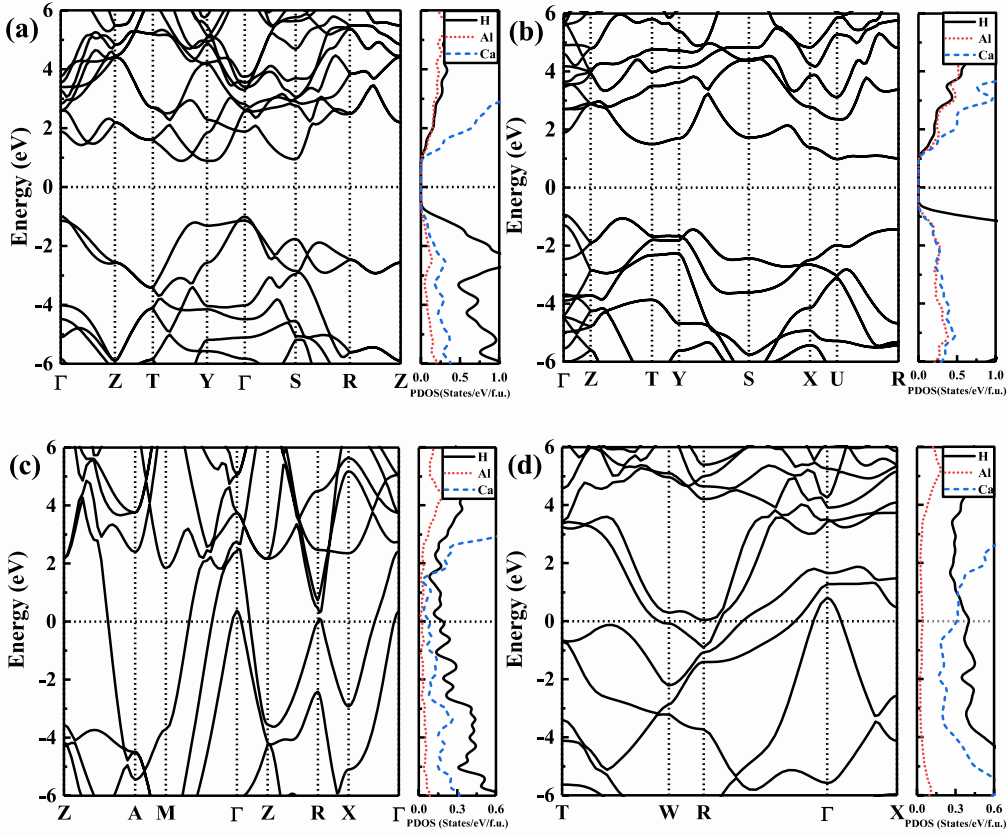

Fig.3. The band structures and partial electronic density of states (PDOS) of (a) $Cmcm$-$CaAlH_5$ at 50 GPa and (b) $Pnnm$-$CaAl_2H_8$ at 50 GPa, (c) $P4/mmm$-$CaAlH_7$ at 200 GPa, and (d) $Immm$-$Ca_2AlH_{12}$ at 250 GPa.

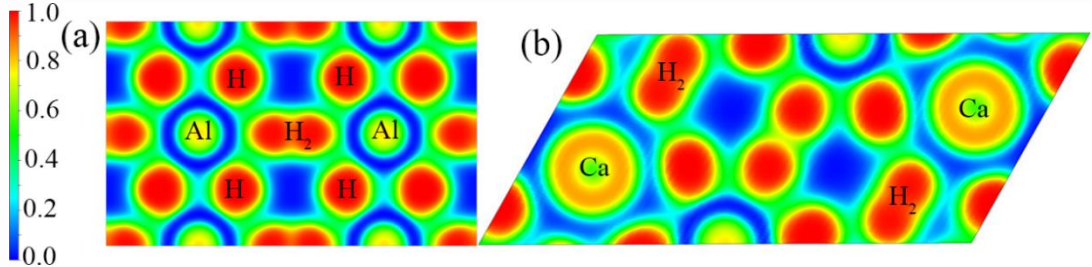

Fig.4. Electronic localization function (ELF) of (a) $P4/mmm$-$CaAlH_7$ at 200 GPa and (b) $Immm$-$Ca_2AlH_{12}$.

To explore the bonding information of these compounds, we calculated the electron localization function (ELF) of $P4/mmm$-$CaAlH_7$ and $Immm$-$Ca_2AlH_{12}$, as shown in Fig. 4. For $P4/mmm$-$CaAlH_7$ and $Immm$-$Ca_2AlH_{12}$, there are no localized electrons between Ca/Al and H, indicating that between Ca/Al and H are pure ionic bonds. The ELF values between the nearest hydrogen atoms of *P4/mmm*-CaAlH$_7$ and *Immm*-Ca$_2$AlH$_{12}$ are 0.93 and 0.95, respectively, indicating that there is a relatively strong covalent bond between the H atoms. Their charge transfer is different, so they correspond to different ELF values. Moreover, the transfer of charge also affects the distance between H molecules.

In order to further understand their bonding information, we calculated the bader charges of CaAlH$_7$ at 50-200 GPa and Ca$_2$AlH$_{12}$ at 250 GPa. The bader charge analysis in Table II shows that the charge is transferred from Ca and Al to H$_2$ and isolated H atoms. In CaAlH$_7$, Ca and Al lose charges greater than 1.05$e$ and 2.42$e$, respectively. The gain of each isolated H atom is about 0.48-0.5$e$ to form H$^-$ ions, and within 50-200 GPa, the electron gains of the H$_2$ unit electrons are 0.52$e$, 0.44$e$, 0.47$e$, 0.5$e$, respectively. At 250 GPa, Ca and Al lose charges greater than 0.88$e$ and 2.39$e$ in Ca$_2$AlH$_{12}$, respectively. Each isolated H atom has a gain of 0.34-0.35$e$, forming an H$^-$ ion, and the electron gain of the H$_2$ unit is 0.17$e$ in Ca$_2$AlH$_{12}$. It can be clearly seen that Ca is a better electron donor than Na under corresponding pressure[26]. Furthermore, the maximum distance between H atoms in the H$_2$ unit in CaAlH$_7$ is 0.93 Å, and the distance between H atoms in the H$_2$ molecule in Ca$_2$AlH$_{12}$ is 0.84 Å, which is obviously longer than the H-H bond of free H$_2$, which is 0.74 Å. It is also longer than the H$_2$ bond length in the Na-Al-H system. We think this is the transfer of Ca and Al electrons to H$_2$, and these extra electrons occupy the antibonding orbital of H-H, thus weakening the strength of H-H bond, thereby increasing the distance of H-H. Especially for CaAlH$_7$, the length of H-H bond increases monotonically with the number of extra electrons within 100-200 GPa. Furthermore, compared with sodium, calcium can provide more electrons to occupy the antibonding orbital of H-H bond, thus weakening the covalent bond between quasi-hydrogen molecules to a greater extent, resulting in the emergence of more quasi-atomic hydrogen, which makes the electrons of hydrogen in the Fermi level of CaAlH$_7$ more than NaAlH$_7$ and NaAlH$_8$[26]. This means that it is possible to be beneficial to the superconducting properties for the CaAlH$_7$.

TABLE II. Bader analysis for CaAlH$_7$ at 50, 100, 150, and 200 GPa, and Ca$_2$AlH$_{12}$ at 250 GPa.

|  | 50 GPa | 100 GPa | 150 GPa | 200 GPa | 250 GPa |
|---|---|---|---|---|---|
| $\delta(e)$ | | CaAlH$_7$ | | | Ca$_2$AlH$_{12}$ |
| H | -0.49~-0.5 | -0.48~-0.49 | -0.49~-0.5 | -0.48~-0.49 | -0.34~-0.35 |
| H$_2$ | -0.52 | -0.44 | -0.47 | -0.5 | -0.17 |
| Ca | 1.05 | 1.05 | 1.04 | 0.99 | 0.88 |
| Al | 2.42 | 2.37 | 2.43 | 2.42 | 2.39 |

### C. Phonon dispersion and superconductivity properties of ternary hydrides

Phonon dispersion curves, projected phonon density of states (PHDOS), Eliashberg spectral function $\alpha^2F(\omega)$, and electronic phonon integrals $\lambda(\omega)$ for *P4/mmm*-CaAlH$_7$ and *Immm*-Ca$_2$AlH$_{12}$ are shown in Fig.5. There are no imaginary frequencies throughout the Brillouin zone for all the ternary hydrides, suggesting the dynamical stability. The minimum pressure for dynamical stable of *P4/mmm*-CaAlH$_7$ is 50 GPa, but the minimum pressure for dynamical stability of *Immm*-Ca$_2$AlH$_{12}$ is 250 GPa. For these stable structures, their contributions to the phonon spectrum and PHDOS are also significantly different due to the different atomic masses. The low frequency vibration modes (0-15 THz) are mainly related to Ca and Al atoms, and the high frequency region is mainly related to the H atom vibration modes, as shown in the Fig.5. The peak of the Eliashberg spectral function $\alpha^2F(\omega)$ and the larger rising region of the electron-phonon integral $\lambda(\omega)$ appear in the high-frequency region associated with the vibrational modes of the H atom, indicating that the hydrogen atoms play a dominant role in the electron-phonon interaction. Importantly, we find that many soft phonon modes emerge in the mid-frequency region of the phonon dispersion curve, as shown in Fig.5. At 200 GPa, $P4/mmm$-CaAlH$_7$, along the Z-A direction, and points A and X appear significantly softening of phonon modes. When the pressure is 50 GPa, the softening of these optical branches will be more obvious. The spectral function $\alpha^2F(\omega)$ and EPC integral $\lambda(\omega)$ are also shown in Fig.5. At 50 GPa, it is found that the low-frequency and high-frequency EPC contribute for 39% and 61% of the total EPC, respectively. At 200 GPa, the proportions are 22% and 78%, respectively. In other words, at 50 and 200 GPa, the derivatization frequencies of H contributed 61% and 78% of the total, respectively, which is relatively common in hydrides at high pressure. In addition, the superconducting transition temperatures were calculated by solving the Allen-Dynes modified McMillan equation[38]:

$$T_c = \frac{\omega_{\log}}{1.2} \exp\left[-\frac{1.04(1+\lambda)}{\lambda - \mu^*(1+0.62\lambda)}\right]$$

Coulomb potential $\mu^*$=0.1-0.13, for $\lambda$<1.5, this equation can provide high precision $T_c$ value. As shown in Table III, at 50 GPa, the $T_c$ of $P4/mmm$-CaAlH$_7$ is 71 K, the electron-phonon coupling coefficient $\lambda$ is 1.28, the logarithmic average phonon frequency $\omega_{log}$ is 658 K. With the pressure increases, the phonons tend to exhibit hardening, the $\omega_{log}$ slowly increases, and the EPC parameter $\lambda$ decreases significantly, so the $T_c$ of CaAlH$_7$ decreases significantly.

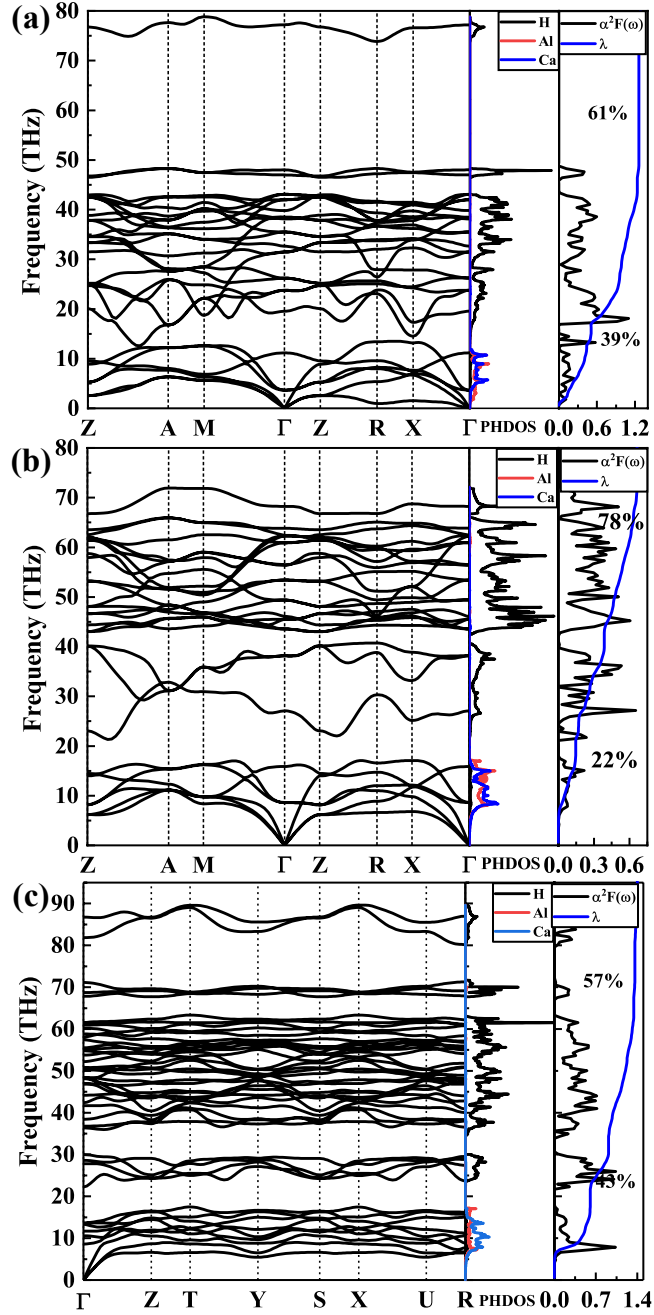

Fig.5. Calculated phonon dispersion curves, projected phonon density of states (PHDOS), and Eliashberg spectral function $\alpha^2F(\omega)$ together with the electron-phonon integral $\lambda(\omega)$ for (a) $P4/mmm$-CaAlH$_7$ at 50 GPa, (b) $P4/mmm$-CaAlH$_7$ at 200 GPa, (c) $Immm$-Ca$_2$AlH$_{12}$ at 250 GPa.

TABLE III. The calculated EPC parameter $\lambda$, logarithmic average phonon frequency $\omega_{log}$ (K), electronic density of states at Fermi level N($\varepsilon_f$) (states/spin/Ry/$f.u.$), and superconducting transition temperatures $T_c$ (K) with $\mu^*$=0.1–0.13 at corresponding pressures P (GPa).

| Structure | P (GPa) | $\lambda$ | $\omega_{log}$ (K) | N($\varepsilon_f$) (states/spin/Ry/$f.u.$) | $T_c$ (K) |
|---|---|---|---|---|---|
| $P4/mmm$-CaAlH$_7$ | 50 | 1.28 | 658 | 3.9 | 62-71 |
|  | 100 | 0.82 | 1166 | 3.2 | 49-61 |
|  | 150 | 0.72 | 1367 | 2.9 | 41-53 |
|  | 200 | 0.66 | 1498 | 2.8 | 35-47 |
| $Immm$-Ca$_2$AlH$_{12}$ | 250 | 1.4 | 979 | 6.8 | 105-118 |

## IV. CONCLUSIONS

In summary, we comprehensively investigated the structural phase diagram, electronic properties, lattice dynamics and superconductivity of the Ca-Al-H ternary system under pressure. We found stable $P4/mmm$-$CaAlH_7$ and metastable $Immm$-$Ca_2AlH_{12}$ under high pressure, the superconducting transition temperature of $CaAlH_7$ can reach 71 K at 50 GPa, and the superconducting transition temperature of $Ca_2AlH_{12}$ can reach 118 K at 250 GPa. The phonon softening along the Z-A direction and at point X have positive contribution to superconductivity of $CaAlH_7$. In addition, we proposed several possible synthetic routes for Ca-Al-H compounds, which will contribute to subsequent experimental studies. These results also provide an important reference for the exploration of other superconducting hydrides.

# Supplementary information for

## "Phase diagrams and superconductivity of ternary Ca-Al-H compounds under high pressure"

Ming Xu[1], Defang Duan[2], Wendi Zhao[1], Decheng An[2] Hao Song[1,*] and Tian Cui[1,2]†

[1]*Institute of High Pressure Physics, School of Physical Science and Technology, Ningbo University, Ningbo 315211, China*

[2]*State Key Laboratory of Superhard Materials, College of Physics, Jilin University, Changchun 130012, China*

*Corresponding author: songhao@nbu.edu.cn
†Corresponding author: cuitian@nbu.edu.cn

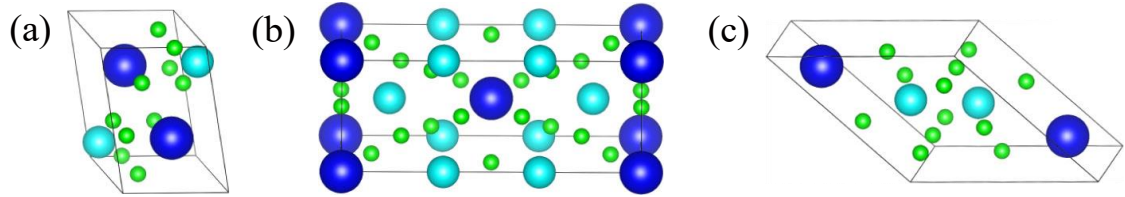

Fig. S1. The crystal structures of (a) *Cmcm*-CaAlH$_5$ (b) *Pnnm*-CaAl$_2$H$_8$ (c) *P*2$_1$/*m*-CaAlH$_5$. Among them, calcium atoms are dark blue, aluminum atoms are light blue, and hydrogen atoms are green.

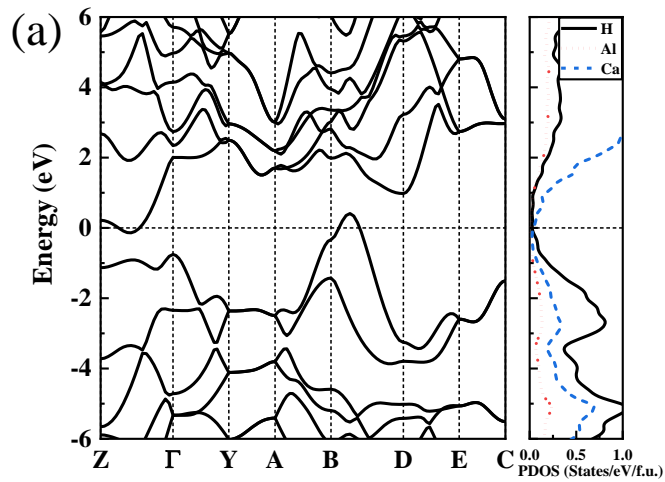

Fig. S2. The band structures and partial electronic density of states (PDOS) of *P*2$_1$/*m*-CaAlH$_5$ at 200 GPa using the GGA-PBE functionals.

i

Table S1. Structural information of predicted hydrides.

| Structure | Lattice parameters(Å) | Atomic coordinates | | | | Sites |
|---|---|---|---|---|---|---|
| P4/mmm CaAlH$_7$ 200 GPa | a=b=2.6015 c=4.2674 α=β=γ=90° | H1 H2 H3 Al Ca | 0.000000 0.000000 0.500000 0.000000 -0.500000 | 1.000000 0.500000 0.500000 0.000000 0.500000 | -0.889920 -0.288020 -0.500000 -0.500000 0.000000 | 2g 4i 1d 1b 1c |
| P2$_1$/m CaAlH$_5$ 200 GPa | a= 5.0112 b= 3.7364 c= 3.2078 α=γ=90° β= 63.9052° | H1 H2 H3 Al Ca | 0.495540 -0.115550 0.194180 0.662960 0.186540 | 0.496650 0.454200 0.250000 0.750000 0.750000 | 0.206350 0.244450 1.210440 0.377780 0.209030 | 4f 4f 2e 2e 2e |
| Cmcm CaAlH$_5$ 50 GPa | a= 3.8830 b= 3.4174 c= 8.8034 α=β=γ=90° | H1 H2 H3 Al Ca | -0.114290 0.748210 1.000000 0.500000 0.000000 | 0.211810 -0.505780 0.000000 0.000000 -0.500000 | 0.800710 0.500000 0.601300 0.839420 1.000000 | 8h 4g 4e 4f 4c |
| Ca$_2$AlH$_{12}$ Immm 200 GPa | a= 8.0124 b= 4.4908 c= 2.6524 α=β=γ=90° | H1 H2 H3 H4 Al Ca | 0.000000 0.343450 0.186400 0.000000 0.000000 0.342920 | -0.398440 -0.716860 -0.908070 -0.805120 0.000000 1.000000 | -1.500000 -1.000000 -1.000000 -2.000000 -0.500000 -1.500000 | 4h 8n 8n 4g 2c 4f |
| CaAl$_2$H$_8$ Pnnm 50 GPa | a= 3.8830 b= 3.4174 c= 8.8034 α=β=γ=90° | H1 H2 H3 Al Ca | -0.114290 0.748210 1.000000 0.500000 0.000000 | 0.211810 -0.505780 0.000000 0.000000 -0.500000 | 0.800710 0.500000 0.601300 0.839420 1.000000 | 8h 4g 4e 4f 4c |